\documentclass[12pt,]{article}
      \usepackage{graphicx}
      \usepackage[cp1251]{inputenc} 
\begin{document}

 \noindent {\footnotesize\it Astronomy Letters, 2010 Vol. 36, No. 9, pp. 634--644.}

 \noindent
 \begin{tabular}{llllllllllllllllllllllllllllllllllllllllllllll}
 & & & & & & & & & & & & & & & & & & & & & & & & & & & & & & & & & & & & &  \\\hline\hline
 \end{tabular}

 \vskip 0.5cm
 \centerline {\large\bf Parameters of the Local Warp of the Stellar-Gaseous}
 \centerline {\large\bf Galactic Disk from the Kinematics of Tycho-2 }
 \centerline {\large\bf  Nearby Red Giant Clump Stars}
 \bigskip
 \centerline {V.V. Bobylev}
 \bigskip
 \centerline {\small\it
 Pulkovo Astronomical Observatory, Russian Academy of Sciences,
 St-Petersburg}
 \bigskip

{\bf Abstract--}We analyze the three-dimensional kinematics of
about 82000 Tycho-2 stars belonging to the red giant clump (RGC).
First, based on all of the currently available data, we have
determined new, most probable components of the residual rotation
vector of the optical realization of the ICRS/HIPPARCOS system
relative to an inertial frame of reference,
 $(\omega_x,\omega_y,\omega_z)=
(-0.11,0.24,-0.52)\pm(0.14,0.10,0.16)$ mas yr$^{-1}$. The stellar
proper motions in the form $\mu_\alpha cos \delta$ have then be
corrected by applying the correction $\omega_z = -0.52$ mas
yr$^{-1}$. We show that, apart from their involvement in the
general Galactic rotation described by the Oort constants
 $A= 15.82\pm0.21$ km s$^{-1}$ kpc$^{-1}$ and
 $B=-10.87\pm0.15$ km s$^{-1}$ kpc$^{-1}$, the
RGC stars have kinematic peculiarities in the Galactic $yz$ plane
related to the kinematics of the warped stellar-gaseous Galactic
disk. We show that the parameters of the linear Ogorodnikov–Milne
model that describe the kinematics of RGC stars in the $zx$ plane
do not differ significantly from zero. The situation in the $yz$
plane is different. For example, the component of the solid-body
rotation vector of the local solar neighborhood around the
Galactic $x$ axis is
 $M_{\scriptscriptstyle32}^{\scriptscriptstyle-} = -2.6\pm0.2$ km s$^{-1}$ kpc$^{-1}$.
Two parameters of the deformation tensor in this plane, namely
 $M_{\scriptscriptstyle23}^{\scriptscriptstyle+} = 1.0\pm0.2$ km s$^{-1}$ kpc$^{-1}$ and
 $M_{\scriptscriptstyle33}-M_{\scriptscriptstyle22}= -1.3\pm0.4$ km s$^{-1}$ kpc$^{-1}$,
also differ significantly from zero. On the whole, the kinematics
of the warped stellar-gaseous Galactic disk in the local solar
neighborhood can be described as a rotation around the Galactic
$x$ axis (close to the line of nodes of this structure) with an
angular velocity
 $(-3.1\pm0.5)\leq\Omega_W\leq(-4.4\pm0.5)$ km s$^{-1}$ kpc$^{-1}$.


\subsection*{INTRODUCTION}

Analysis of the large-scale structure of neutral hydrogen revealed
a warp of the gaseous disk in the Galaxy (Westerhout 1957; Burton
1988). The results of studying this structure based on currently
available data on the HI and HII distributions are presented in
Kalberla and Dedes (2008) and Cersosimo et al. (2009),
respectively. This structure is revealed by the spatial
distribution of stars and dust (Drimmel and Spergel 2001), by the
distribution of pulsars in the Galaxy (Yusifov 2004), by HIPPARCOS
OB stars (Miyamoto and Zhu 1998), and by the 2MASS red giant clump
(Momany et al. 2006).

Several models were suggested to explain the nature of the
Galactic disk warp: (1) the interaction between the disk and a
nonspherical dark matter halo (Sparke and Casertano 1988); (2) the
gravitational influence of the Galaxy’s nearest satellites (Bailin
2003); (3) the interaction of the disk with a near-Galaxy flow
formed by high-velocity hydrogen clouds that resulted from mass
exchange between the Galaxy and the Magellanic Clouds (Olano
2004); (4) an intergalactic flow (L\'opez-Corredoira et al. 2002);
and (5) the interaction with the intergalactic magnetic field
(Battaner et al. 1990).

By analyzing nearby stars from HIPPARCOS (1997), Dehnen (1998)
showed that the distribution of their residual velocities in the
$V_y-V_z$ plane agreed satisfactorily with various rotation models
of the warped disk. Miyamoto et al. (1993) and Miyamoto and Zhu
(1998) determined the rotation parameters of the warped
stellar–gaseous disk by analyzing giant stars of various spectral
types and a sample of HIPPARCOS O–B5 stars. Thus, there is
positive experience in solving this problem using data on stars
relatively close to the Sun.

Studying the three-dimensional kinematics of stars requires that
the observational data be free from the systematic errors related
to the referencing of the optical realization of the
ICRS/HIPPARCOS system to the inertial frame of reference specified
by extragalactic sources. The modern standard system of
astronomical coordinates, ICRS (International Celestial Reference
System), is realized by the catalog of positions for 212 compact
extragalactic radio sources uniformly distributed over the entire
sky observed by the radio interferometry technique (Ma et al.
1998). In the optical range, the first realization of the ICRS was
the HIPPARCOS catalog. The application of various methods of
analysis shows that there is a small residual rotation of the
ICRS/HIPPARCOS system relative to the inertial frame of reference
with $\omega_z\approx-0.4\pm0.1$ mas yr$^{-1}$ (Bobylev 2004a,
2004b). At present, there are several new results of comparing
individual programs with the catalogs of the ICRS/HIPPARCOS
system. One of our goals is to determine the most probable
components of the residual rotation vector of the optical
realization of the ICRS/HIPPARCOS system relative to the inertial
frame of reference.

The main goal of this paper is to study the local kinematics of
the warped stellar–gaseous Galactic disk by analyzing the motions
of Tycho-2 stars that belong to the red giant clump (RGC).
Occupying a compact region on the Hertzsprung–Russell diagram,
these stars are a kind of ``standard candles''. Their estimated
photometric distances are known with a mean accuracy of at least
20--30\%. RGC stars are distributed uniformly in the spatial
region and over the celestial sphere, which is an important
property when the three-dimensional spatial motions of stars are
analyzed.

\begin{figure}[p]{
\begin{center}
 \includegraphics[width=80mm]{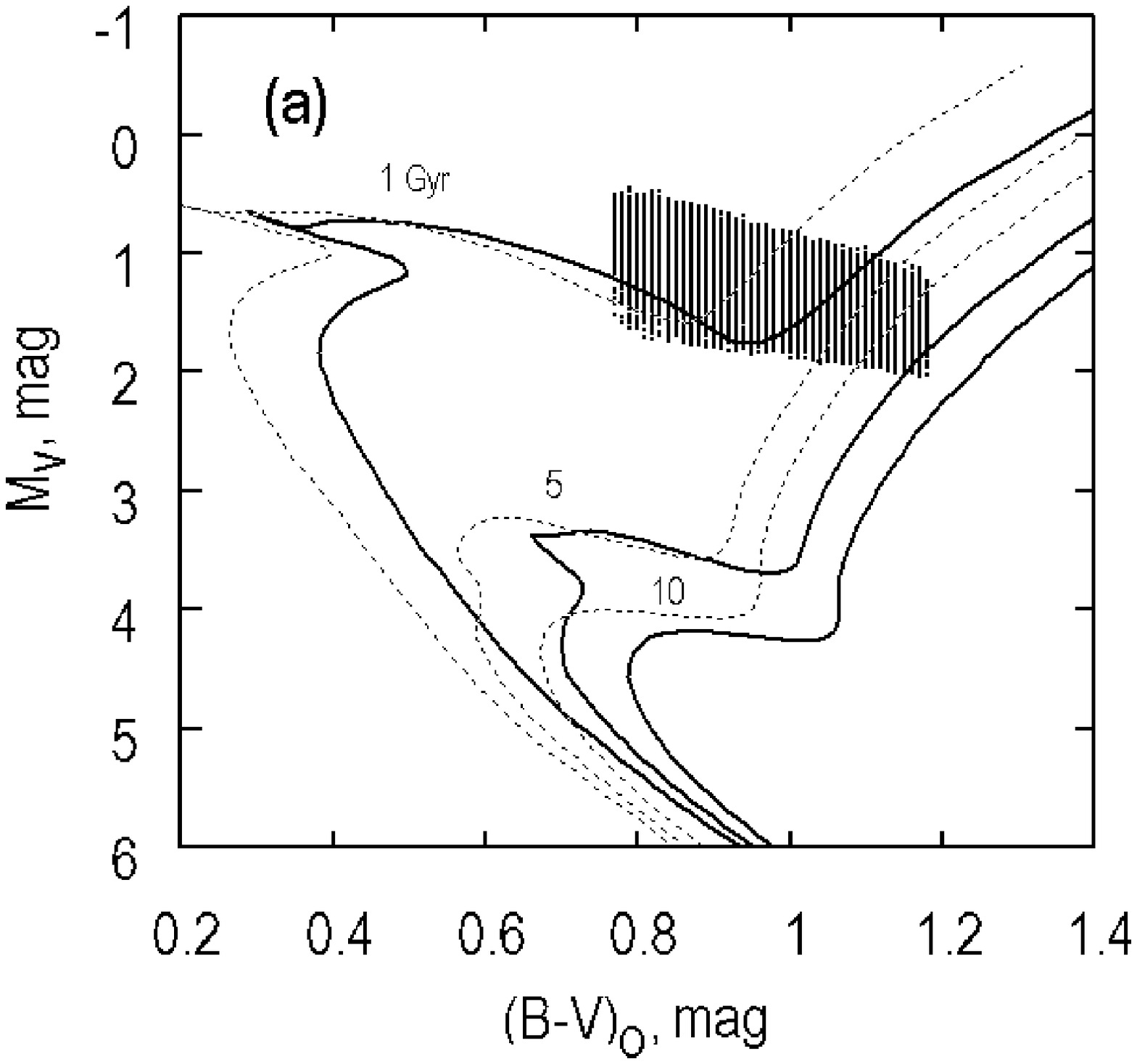}

 \includegraphics[width=80mm]{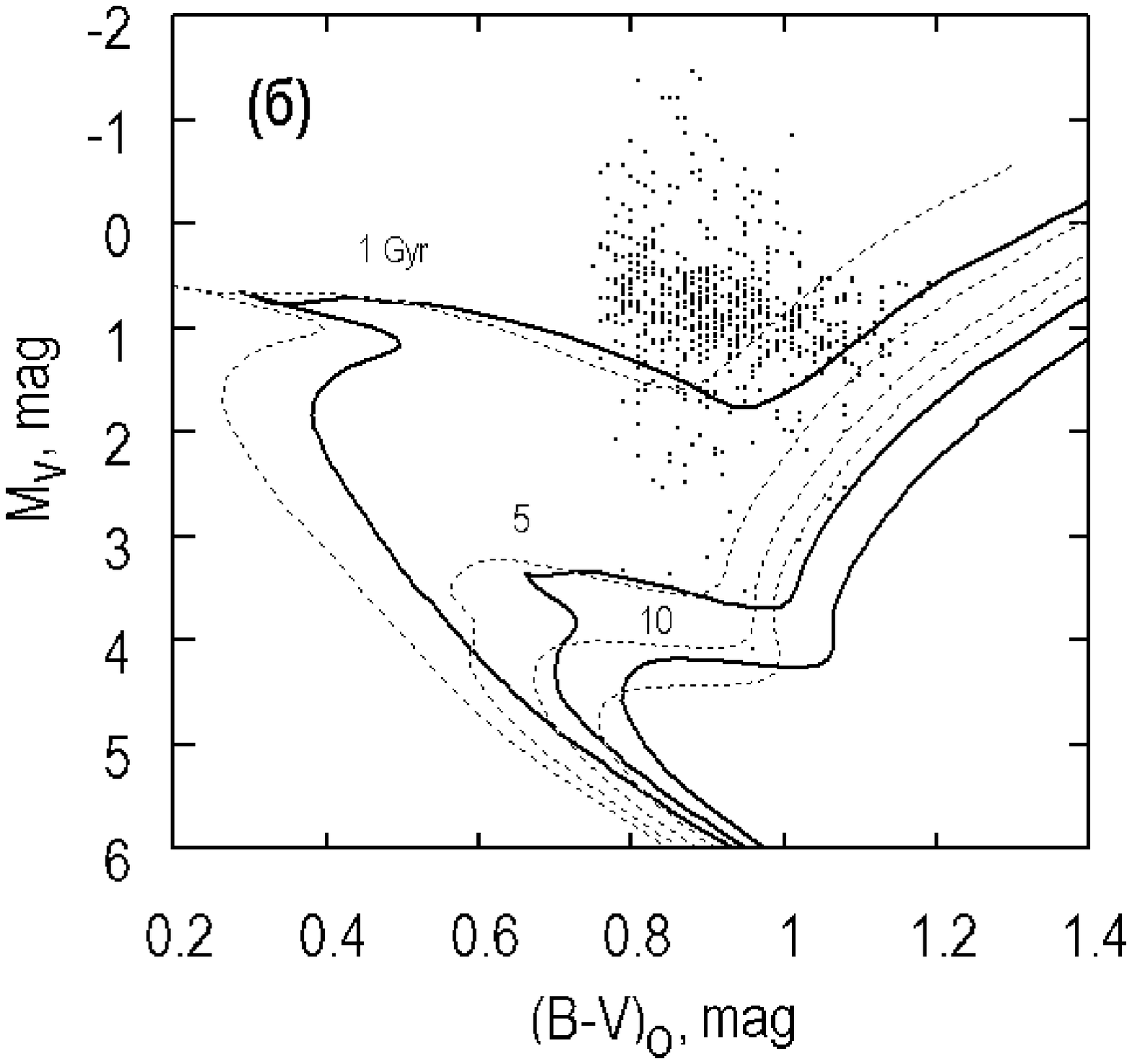}
\end{center}}
Fig. 1. Color–absolute magnitude diagram for 82324 RGC stars from
the range of distances 0.3--1 kpc (a) and the sample of nearby RGC
stars with reliable ($e_\pi/\pi < 10$\%) HIPPARCOS parallaxes (b).
The isochrones (Yi et al. 2003) for ages of 1, 5, and 10 Gyr with
a nearly solar metallicity, $Z=0.02$ (dashed lines), and a higher
metallicity, $Z=0.04$ (solid lines), are plotted.
\end{figure}

\subsection*{DATA}

The characteristic clump, commonly called the red giant clump, on
the Hertzsprung–Russell diagram is formed mostly by giants with
masses from 3$M_\odot$ to 9$M_\odot$. These stars spend the bulk
of their lifetime on the main sequence as B-type stars. At the
core helium burning stage, they evolve toward the RGC almost
without changing their luminosity. At the RGC stage, despite the
slow change in color, the luminosity of such a star remains almost
constant. Therefore, RGC stars are convenient as ``standard
candles'' for distance determinations. The RGC also incorporates
other giants at different evolutionary stages. In addition, owing
to probabilistic selection methods using reduced proper motions,
the RGC sample can be diluted by various stars from adjacent
regions of the Hertzsprung–Russell diagram, both supergiants and
dwarfs. According to the estimates by various authors, the
admixture is small, may be 10--15\% (Rybka 2006; Gontcharov 2008).

We used the list of 97000 RGC stars selected by Gontcharov (2008)
by the dereddened color index and the reduced proper motion based
on Tycho-2 (Hog et al. 2000) and 2MASS (Skrutskie et al. 2006)
data. These stars occupy the region with $0.^m5 < J-Ks < 0^m.9$
and $-1^m<M_{K_s}<-2^m$ on the corresponding diagram. For their
selection using the reduced proper motion, calibration based on
HIPPARCOS stars with the most reliable data was performed. Details
of the procedure are described in Gontcharov (2008). An estimate
of the photometric distance and interstellar extinction is
available for each star.

As was shown by Bobylev et al. (2009), the kinematic parameters of
the Ogorodnikov–Milne model for relatively close ($r<0.2-0.3$ kpc)
and distant ($r>1$ kpc) stars are determined with large errors
when using the photometric distances of RGC stars. Therefore, here
we use 82 324 stars from the range of distances 0.3 kpc $<r<1$
kpc, with the mean distance being $r = 0.57\pm0.17$ kpc. To
eliminate the halo stars from the sample, the following constraint
on the proper motions was applied:
 $\sqrt{(\mu_\alpha\cos\delta)^2+(\mu_\delta)^2}<300$ mas yr$^{-1}$.

As was shown by Bobylev et al. (2009), from a statistical analysis
of the velocity dispersions, we found that about 20\% of the RGC
stars are very young, while an overwhelming majority of the
remaining stars are characterized by the kinematics of thin-disk
stars. We see from Fig. 1, where the color--absolute magnitude
diagram is presented with a grid of isochrones with various ages
and metallicities, that the RGC stars (Fig. 1a) are fairly young.
Figure 1b presents about 650 stars from Gontcharov’s catalog
selected as RGC ones with parallaxes from the HIPPARCOS catalog
(with a parallax error of less than 10\%). Although the segment of
the isochrones near the RGC is very sensitive to metallicity, we
clearly see from Fig. 1b that, first, the RGC stars proper have
ages younger than 1 Gyr and, second, the admixture of dwarfs is
insignificant.

\subsection*{THE OGORODNIKOV–MILNE MODEL}

We use a rectangular Galactic coordinate system with the axes
directed away from the observer toward the Galactic center
$(l$=$0^\circ$, $b$=$0^\circ,$ the $X$ axis or axis 1), along the
Galactic rotation $(l$=$90^\circ$,~$b$=$0^\circ,$ the $Y$ axis or
axis 2), and toward the North Galactic Pole $(b$=$90^\circ,$ the
$Z$ axis or axis 3).

In the linear Ogorodnikov-Milne model (Ogorodnikov, 1965), we use
the notation introduced by Clube (1972, 1973) and used, for
example, by Vityazev and Tsvetkov (2009).

The observed velocity ${\bf V}(r)$ of a star with a heliocentric
radius vector ${\bf r}$, is described, to the terms of the first
order of smallness $r/R_0\ll 1$, by the equation in vector form
 $$
 \displaylines{ \hfill
 {\bf V}(r)={\bf V}_\odot+M{\bf r}+{\bf V'},\hfill\llap(1)\cr}
 $$
where ${\bf V}_\odot(X_\odot,Y_\odot,Z_\odot)$ is the peculiar
velocity of the Sun relative to the centroid of the stars under
consideration; {$\bf V'$} is the residual velocity of the star
(here, the residual stellar velocities are assumed to have a
random distribution); $M$ is the displacement matrix (tensor)
whose components are the partial derivatives of the velocity ${\bf
u}(u_1,u_2,u_3)$ with respect to the distance ${\bf r}(r_1, r_2,
r_3)$, where ${\bf u}={\bf V}(R)-{\bf V}(R_0)$, and $R$ and $R_0$
are the Galactocentric distances of the star and the Sun,
respectively. Then,
 $$ \displaylines{\hfill
 M_{pq}={\left(\frac{\partial u_p} {\partial r_q}\right)}_\circ, \quad (p,q=1,2,3). \hfill\llap(2)\cr
  }
 $$
All nine elements of the matrix $M$ can be determined using the
three components of the observed velocities--the stellar radial
velocities and proper motions. Having only the proper motions, we
can write the conditional equations
 $$ \displaylines{\hfill
 4.74 r \mu_l\cos b= X_\odot\sin l-Y_\odot\cos l+ \hfill\llap(3) \cr\hfill
 +r [-\cos b\cos l\sin l  M_{11}
 -\cos b\sin^2 l M_{12}  
 -\sin b \sin l  M_{13}
 +\cos b\cos^2 l M_{21}+
\hfill\cr\hfill
 +\cos b\sin l\cos l  M_{22} 
 +\sin b\cos l  M_{23} ],\hfill\cr\hfill
 4.74 r \mu_b=X_\odot\cos l\sin b 
 +Y_\odot\sin l\sin b-Z_\odot\cos b+ \hfill\llap{(4)}\cr\hfill
 +r [-\sin b\cos b\cos^2 l M_{11} 
 -\sin b\cos b\sin l \cos l M_{12}- \hfill\cr\hfill
 -\sin^2 b \cos l  M_{13}
 -\sin b\cos b\sin l\cos l M_{21} 
 -\sin b\cos b\sin^2 l  M_{22}
 -\sin^2 b\sin l  M_{23}+          \hfill\cr\hfill
 +\cos^2 b\cos l M_{31}
 +\cos^2 b\sin l M_{32} 
 +\sin b\cos b  M_{33} ],
\hfill }
$$
from which it follows that the terms should be grouped in several
cases. It is useful to divide the matrix $M$ into symmetric,
$M^{\scriptscriptstyle+}$ (local deformation tensor), and
antisymmetric, $M^{\scriptscriptstyle-}$ (rotation tensor), parts:
$$
\displaylines{\hfill
 M_{\scriptstyle pq}^{\scriptscriptstyle+}=
 {1\over 2}\left( \frac{\partial u_{p}}{\partial r_{q}}+
 \frac{\partial u_{q}}{\partial r_{p}}\right)_\circ, \quad 
 M_{\scriptstyle pq}^{\scriptscriptstyle-}=
 {1\over 2}\left(\frac{\partial u_{p}}{\partial r_{q}}-
 \frac{\partial u_{q}}{\partial r_{p}}\right)_\circ, \quad 
 (p,q=1,2,3).\hfill\llap(5) }
 $$
This allows the conditional equations to be written as
$$\displaylines{\hfill 4.74 r \mu_{l}\cos b=
       X_{\odot}\sin l-Y_{\odot}\cos l+\hfill\llap(6)\cr\hfill
 +r[-M_{\scriptscriptstyle32}^{\scriptscriptstyle-}\cos l\sin b
   -M_{\scriptscriptstyle13}^{\scriptscriptstyle-}\sin l\sin b
   +M_{\scriptscriptstyle21}^{\scriptscriptstyle-}\cos b+
\hfill\cr\hfill
   +M_{\scriptscriptstyle12}^{\scriptscriptstyle+}\cos 2l\cos b
   -M_{\scriptscriptstyle13}^{\scriptscriptstyle+}\sin l\sin b 
   +M_{\scriptscriptstyle23}^{\scriptscriptstyle+}\cos l\sin b
  -0.5(M_{\scriptscriptstyle11}^{\scriptscriptstyle+}
  -M_{\scriptscriptstyle22}^{\scriptscriptstyle+})\sin 2l\cos b],
\hfill \cr\hfill
4.74 r \mu_b=
    X_{\odot}\cos l\sin b+Y_{\odot}\sin l\sin b-Z_{\odot}\cos b+\hfill\llap(7)\cr\hfill
 +r[M_{\scriptscriptstyle32}^{\scriptscriptstyle-}\sin l
   -M_{\scriptscriptstyle13}^{\scriptscriptstyle-}\cos l 
-0.5M_{\scriptscriptstyle12}^{\scriptscriptstyle+}\sin 2l\sin 2b
   +M_{\scriptscriptstyle13}^{\scriptscriptstyle+}\cos l\cos 2b+
\hfill\cr\hfill
   +M_{\scriptscriptstyle23}^{\scriptscriptstyle+}\sin l\cos 2b
-0.5(M_{\scriptscriptstyle11}^{\scriptscriptstyle+}
    -M_{\scriptscriptstyle22}^{\scriptscriptstyle+})\cos^2 l\sin 2b 
+0.5(M_{\scriptscriptstyle33}^{\scriptscriptstyle+}
    -M_{\scriptscriptstyle22}^{\scriptscriptstyle+})\sin 2b].
\hfill }
 $$
Equation (6) was obtained from Eq. (3) by adding two terms to its
right-hand side, $0.5M_{\scriptscriptstyle31}$ and
$0.5M_{\scriptscriptstyle32}$, and subtracting them.

The system of equations (6) and (7) then becomes very convenient
for its simultaneous solution. Indeed, as can be seen from Eq.
(6), the two pairs of unknowns to be determined,
$M_{\scriptscriptstyle13}^{\scriptscriptstyle-}$ and
$M_{\scriptscriptstyle13}^{\scriptscriptstyle+}$, along with
$M_{\scriptscriptstyle32}^{\scriptscriptstyle-}$ and
$M_{\scriptscriptstyle23}^{\scriptscriptstyle+}$, have identical
coefficients, $\sin l\sin b$ and $\cos l\sin b,$ respectively. In
this case, the variables cannot be separated and can be found only
from the simultaneous solution of the system of equations (6) and
(7).

In addition, we see that one of the diagonal terms of the local
deformation tensor remains uncertain. Therefore, we determine
differences of the form
$(M_{\scriptscriptstyle11}^{\scriptscriptstyle+}-
 M_{\scriptscriptstyle22}^{\scriptscriptstyle+})$ and
$(M_{\scriptscriptstyle33}^{\scriptscriptstyle+}-
 M_{\scriptscriptstyle22}^{\scriptscriptstyle+})$.

The quantities $M_{\scriptscriptstyle32}^{\scriptscriptstyle-},
 M_{\scriptscriptstyle13}^{\scriptscriptstyle-},
 M_{\scriptscriptstyle12}^{\scriptscriptstyle-},$ are the components of the
solid-body rotation vector of a small solar neighborhood around
the $x,y,z$ axes, respectively. According to our chosen
rectangular coordinate system, the rotations from axis 1 to 2,
from axis 2 to 3, and from axis 3 to 1 are positive.
$M_{\scriptscriptstyle21}^{\scriptscriptstyle-}$ is equivalent to
the Oort constant $B$. Each of the quantities
$M_{\scriptscriptstyle12}^{\scriptscriptstyle+},
  M_{\scriptscriptstyle13}^{\scriptscriptstyle+},
  M_{\scriptscriptstyle23}^{\scriptscriptstyle+}$
describes the deformation in the corresponding plane.
$M_{\scriptscriptstyle12}^{\scriptscriptstyle+}$ is equivalent to
the Oort constant $A$.

The diagonal components of the local deformation tensor
$M_{\scriptscriptstyle11}^{\scriptscriptstyle+},
  M_{\scriptscriptstyle22}^{\scriptscriptstyle+},
  M_{\scriptscriptstyle33}^{\scriptscriptstyle+}$ describe
the general local contraction or expansion of the entire stellar
system (divergence). The system of conditional equations (6) and
(7) contain eleven unknowns to be determined by the least-squares
method. In addition to the simultaneous solution of the system of
equations (6) and (7), here we analyze the results of their
separate solutions. In this case, Eq. (7) remains unchanged, while
Eq. (6) is reduced to the form
 $$\displaylines{\hfill 4.74 r
\mu_{l}\cos b=
       X_{\odot}\sin l-Y_{\odot}\cos l+\hfill\llap(8)
\cr\hfill
 +r [M_{\scriptscriptstyle23}\cos l\sin b
   -M_{\scriptscriptstyle13}\sin l\sin b 
   +M_{\scriptscriptstyle21}^{\scriptscriptstyle-}\cos b
   +M_{\scriptscriptstyle12}^{\scriptscriptstyle+}\cos 2l\cos b-\hfill\cr\hfill
  -0.5(M_{\scriptscriptstyle11}^{\scriptscriptstyle+}
  -M_{\scriptscriptstyle22}^{\scriptscriptstyle+})\sin 2l\cos b ],\hfill \cr
 }
 $$
where there are only seven independent variables.

\subsection*{INERTIALITY OF THE OPTICAL ICRS/HIPPARCOS SYSTEM}

Table 1 presents all of the currently known results of comparing
individual programs with the catalogs of the ICRS/HIPPARCOS
system.

First, we will briefly describe the results, most of which were
used by Kovalevsky et al. (1997) to calibrate the HIPPARCOS
catalog and to estimate its residual rotation relative to the
system of extragalactic sources and by Bobylev (2004b) to solve
this problem.

(1) The NPM1 solution. We used the result of comparing the stellar
proper motions from the NPM1 (Klemola et al. 1994) and HIPPARCOS
catalogs performed by the Heidelberg group (Kovalevsky et al.
1997). This solution was obtained in the range of magnitudes
$10^m.5–11^m.5,$ where (Fig. 1 in Platais et al. 1998a) the
HIPPARCOS–NPM1 stellar proper motion differences have a
``horizontal'' pattern near zero. In our opinion, the NPM1 proper
motions in this magnitude range are free from the influence of the
magnitude equation, which is significant in this catalog, to the
greatest extent.

{
\begin{table}[t]                                                
\caption[]{\small Components of the residual rotation vector of
the optical realization of the ICRS/HIPPARCOS system relative to
the inertial frame of reference
 }
\begin{center}
\begin{tabular}{|l|c|c|c|c|c|c|c|c|c|c|c|c|c|}\hline

 Method   & $P_x/P_y/P_z$ & $N_\star$ & $N_{\hbox{\tiny area}}$
 & $\omega_x,$~mas yr$^{-1}$
 & $\omega_y,$~mas yr$^{-1}$
 & $\omega_z,$~mas yr$^{-1}$ \\\hline

 NPM1    & 10/5/16  &   2616 &  899 & $-0.76\pm0.25$ & $+0.17\pm0.20$ & $-0.85\pm0.20$ \\
 NPM2    &   (*)    &   3519 &  347 & $-0.11\pm0.20$ & $-0.19\pm0.20$ & $-0.75\pm0.28$ \\
 SPM2    & 22/9/28  &   9356 &  156 & $+0.10\pm0.17$ & $+0.48\pm0.14$ & $-0.17\pm0.15$ \\
 Kiev    &  1/1/1   &    415 &  154 & $-0.27\pm0.80$ & $+0.15\pm0.60$ & $-1.07\pm0.80$ \\
 Potsdam &  2/1/3   &    256 &   24 & $+0.22\pm0.52$ & $+0.43\pm0.50$ & $+0.13\pm0.48$ \\
 Bonn    &  5/3/6   &     88 &   13 & $+0.16\pm0.34$ & $-0.32\pm0.25$ & $+0.17\pm0.33$ \\
 HST     & 0.1/0.1/0.05 & 78 &      & $-1.60\pm2.87$ & $-1.92\pm1.54$ & $+2.26\pm3.42$ \\
 EOP     &  8/2/--- &        &      & $-0.93\pm0.28$ & $-0.32\pm0.28$ & --- \\
 PUL2    &  3/1/4   &   1004 &  147 & $-0.98\pm0.47$ & $-0.03\pm0.38$ & $-1.66\pm0.42$ \\
 XPM     & 28/9/33  & $1\times10^6$ & 1431 & $-0.06\pm0.15$ & $+0.17\pm0.14$ & $-0.84\pm0.14$ \\
 VLBI-07 &  6/1/5   &     46 &      & $-0.55\pm0.34$ & $-0.02\pm0.36$ & $+0.41\pm0.37$ \\
 Minor Pl & 25/5/6  & 116  &      & $+0.12\pm0.08$ & $+0.66\pm0.09$ & $-0.56\pm0.16$ \\\hline
 Mean 1 &        &        &      & $-0.22\pm0.19$ & $+0.14\pm0.10$ & $-0.49\pm0.23$ \\
 Mean 2 &        &        &      & $-0.11\pm0.14$ & $+0.24\pm0.10$ & $-0.52\pm0.16$ \\\hline

\end{tabular}
\end{center}
 \small\baselineskip=1.0ex\protect
 Note. (*)—the NPM2 solution is not used, $N_\star$ is the number
of stars/asteroids, $N_{area}$ is the number of areas on the
celestial sphere, mean 1 is a simple mean (without HST), mean 2 is
a weighted mean.

\vskip6mm
\end{table}
}

(2) The NPM2 solution. The stellar proper motions from the NPM2
and HIPPARCOS catalogs were compared by Zhu (2003). However, there
are no images of galaxies on NPM2 photographic plates and, hence,
these proper motions are relative. Therefore, we do not use this
solution to derive the mean values of
$\omega_x,\omega_y,\omega_z$. It is given in Table 1 to emphasize
its similarity to the NPM1 solution.

(3) The SPM2 solution. The stellar proper motions from the SPM2
(Platais et al. 1998b) and HIPPARCOS catalogs were compared by Zhu
(2001).

(4) The PUL2 solution. The parameters $\omega_x,\omega_y,\omega_z$
were found by comparing the Pulkovo PUL2 photographic catalog
(Bobylev et al. 2004) and HIPPARCOS.

(5) The KIEV solution. The stellar proper motions from the GPM1
(Rybka and Yatsenko 1997) and HIPPARCOS catalogs were compared by
Kislyuk et al. (1997).

(6) The POTSDAM solution. The parameters
$\omega_x,\omega_y,\omega_z$ of the Potsdam program were taken
from Hirte et al. (1996).

(7) The BONN solution. The results of the Bonn program are
presented in Geffert et al. (1997) and Tucholke et al. (1997).

(8) The EOP solution. The results of the analysis of Earth
orientation parameters (EOP) were taken from Vondr\'ak et al.
(1997). Only two rotation parameters, $\omega_x,$ and $\omega_y$
are determined in this method.

(9) The HST solution. The results of stellar observations with the
Hubble Space Telescope (HST) were taken from Hemenway et al.
(1997).

Now, we will point out several new results that have not been used
previously to solve this problem.

(10) The XPM solution. The XPM catalog (Fedorov et al., 2009)
contains absolute proper motions for about 275 million stars
fainter than 12m derived by comparing their positions in the 2MASS
and USNO-A2.0 (Monet 1998) catalogs. The absolutization was made
using about 1.5 million galaxies from the 2MASS catalog of
extended sources. Thus, the XPM catalog is an independent
realization of the inertial frame of reference. The stellar proper
motions from the XPM and UCAC2 (Zacharias et al. 2004) catalogs
were compared by Bobylev et al. (2010). The parameters
$\omega_x,\omega_y,\omega_z$ were calculated using about 1 million
stars. Among all of the programs listed in Table 1, the XPM
solution is unique in that it was obtained from the differences of
stars covering the entire sky almost completely, except for the
zone $\delta > 54^\circ,$ where there are no UCAC2 stars.

(11) The VLBI-07 solution. Boboltz et al. (2007) analyzed the
positions and proper motions of 46 radio stars and obtained new
parameters of the mutual orientation of the optical realization
(HIPPARCOS) and the radio system.

(12) The ``MINOR PLANETS'' solution. Chernetenko (2008) estimated
the rotation parameters of the HIPPARCOS system relative to the
DE403 and DE405 coordinate systems of ephemerides by analyzing a
long-term series of asteroid observations. This result suggests
that either the dynamical DE403 and DE405 theories need to be
improved or the HIPPARCOS system needs to be corrected. We reduced
the weight of this solution by half because of the possible
contribution from the inaccuracy of the dynamical DE403 and DE405
theories.

The weight of each of the comparison catalogs was taken to be
inversely proportional to the square of the random error
$e_\omega$ in the corresponding quantities
$\omega_x,\omega_y,\omega_z$ and was calculated from the formula
$$
\displaylines{\hfill
 P_i={e_{kiev}}^2/{{e_i}^2}, \quad i=1,...,11.\hfill\llap(9)
 }
$$
Not all of the authors use the equations to determine
$\omega_x,\omega_y,\omega_z$ in the form in which they were
suggested by Lindegren and Kovalevsky (1995):
$$
\displaylines{\hfill
 \Delta\mu_\alpha\cos\delta= \omega_x\cos\alpha\sin\delta+
 \omega_y\sin\alpha\sin\delta- \omega_z\cos\delta,\hfill\llap(10)\cr\hfill
 \Delta\mu_\delta=-\omega_x\sin\alpha+ \omega_y\cos\alpha,
 \hfill\llap(11)
 }
$$
where the catalog--HIPPARCOS differences are on the left-hand
sides of the equations. Therefore, in several cases, the signs of
the quoted quantities were reduced to the necessary uniform form
(Zhu 2001, 2003; Boboltz et al. 2007).

The last rows of Table 1 give Mean 1 calculated as a simple mean
and Mean 2 that was calculated as a weighted mean and is the main
result of our analysis.

Denote the components of the rotation vector around the
rectangular equatorial axes by $\omega_x,\omega_y,\omega_z$; then,
$$
\displaylines{\hfill
           \pmatrix{
           \Omega_x\cr
           \Omega_y\cr
           \Omega_z\cr
           }=
{\bf G}\pmatrix {
 M_{\scriptscriptstyle32}^{\scriptscriptstyle-}\cr
 M_{\scriptscriptstyle13}^{\scriptscriptstyle-}\cr
 M_{\scriptscriptstyle21}^{\scriptscriptstyle-}\cr
 }
   +
4.74\pmatrix{ \omega_x\cr
                   \omega_y\cr
                  -\omega_z\cr},\hfill\llap(12) \cr }
$$
where
$$
\displaylines{
  \hfill
           {\bf G}=
\pmatrix{
 -0.0548 & +0.4941 & -0.8677 \cr
 -0.8734 & -0.4448 & -0.1981 \cr
 -0.4838 & +0.7470 & +0.4560 \cr
 }
 \hfill\llap(13) \cr
}
$$
is the well-known transformation matrix between the unit vectors
of the Galactic and equatorial coordinate systems. From Eqs. (9)
and (10), it is easy to see the relationship between
$M_{\scriptscriptstyle13}^{\scriptscriptstyle-}$ and $\omega_z$.
Assuming the components in the mean-2 solution to be zero, $$
\displaylines{\hfill
           \pmatrix{
           \Omega_x\cr
           \Omega_y\cr
           \Omega_z\cr
           }=
{\bf G}\pmatrix {
 M_{\scriptscriptstyle32}^{\scriptscriptstyle-}\cr
 M_{\scriptscriptstyle13}^{\scriptscriptstyle-}\cr
 M_{\scriptscriptstyle21}^{\scriptscriptstyle-}\cr
 }
   +
4.74\pmatrix{ 0\cr
              0\cr
              0.52\cr},\hfill\llap(14) \cr }
$$
for the case where the left- and right-hand sides are expressed in
km s$^{-1}$ kpc$^{-1}$.

\subsection*{KINEMATICS OF THE DISK WARP}

The results of determining the kinematic parameters of the
Ogorodnikov–Milne model using a sample of RGC stars derived by
simultaneously solving the system of equations (6) and (7) are
presented in Table 2. The second column gives the solution
obtained without applying any corrections to the input data; the
third column gives the solution for the case where the stellar
proper motions in the form $\mu_\alpha\cos\delta$  were corrected
by applying the correction $\omega_z=-0.52$ mas yr$^{-1}$ (14)
using Eq. (10).

The parameters derived by separately solving Eqs. (8) and (7) are
presented in Table 3.

As can be seen from Table 2, applying the correction affected only
three components of the rotation tensor: insignificantly
$M_{\scriptscriptstyle21}^{\scriptscriptstyle-}$, noticeably
$M_{\scriptscriptstyle32}^{\scriptscriptstyle-}$, and most
strongly $M_{\scriptscriptstyle13}^{\scriptscriptstyle-}$, which
is explained by the structure of the matrix G (13). Since the
components $M_{\scriptscriptstyle13}^{\scriptscriptstyle-}$ and
 $M_{\scriptscriptstyle13}^{\scriptscriptstyle+}$ found (the third column of Table 2) do
not differ significantly from zero, they may be set equal to zero.

Suppose that the values of
$M_{\scriptscriptstyle21}^{\scriptscriptstyle-}$ and
$M_{\scriptscriptstyle21}^{\scriptscriptstyle+}$ (the Oort
constants) found describe only the rotation around the Galactic
$z$ axis, while the motion in the $yz$ plane is independent. Let
us now consider the displacement tensor $M_W$ that describes the
kinematics in the $yz$ plane:
$$ \displaylines{\hfill
 M_W=\pmatrix{
 M_{\scriptscriptstyle22}& M_{\scriptscriptstyle23}\cr
 M_{\scriptscriptstyle32}&
 M_{\scriptscriptstyle33}\cr}=
   \pmatrix{
 {\strut \displaystyle\partial u_2}\over{\displaystyle\partial r_2}& {\strut \displaystyle\partial u_2}\over{\displaystyle\partial r_3}\cr
 {\strut \displaystyle\partial u_3}\over{\displaystyle\partial r_2}& {\strut \displaystyle\partial u_3}\over{\displaystyle\partial r_3}\cr}
 .\hfill\llap(15)
}
$$
As has already been noted, when using only the stellar proper
motions, we can determine only the difference
$(M_{\scriptscriptstyle33}^{\scriptscriptstyle+}-
 M_{\scriptscriptstyle22}^{\scriptscriptstyle+})$.
 The identity $(M_{\scriptscriptstyle33}^{\scriptscriptstyle+}-
 M_{\scriptscriptstyle22}^{\scriptscriptstyle+})\equiv
 (M_{\scriptscriptstyle33}-M_{\scriptscriptstyle22})$ is valid for
the diagonal elements. As can be seen from Table 2, the value of
this quantity differs significantly from zero.

Three cases are possible when analyzing tensor (15):

 1) $M_{\scriptscriptstyle22}\neq0$, $M_{\scriptscriptstyle33}=0$;

 2) $M_{\scriptscriptstyle22}=0$, $M_{\scriptscriptstyle33}\neq0$;

 3) $M_{\scriptscriptstyle22}=0$, $M_{\scriptscriptstyle33}=0$.

For the completeness of the picture, note that case 4 is also
possible: $M_{\scriptscriptstyle22}\neq0$,
$M_{\scriptscriptstyle33}\neq0$; since this requires data on the
stellar radial velocities, this case is not considered here.

Consider case 1, $M_{\scriptscriptstyle33}=0,$, using the data
from the third column of Table 2,
$M_{\scriptscriptstyle22}=1.3\pm0.4$ km s$^{-1}$ kpc$^{-1}$. The
components of the displacement tensor $M_W$, the symmetric
deformation tensor $M_W^{\scriptscriptstyle+}$, and the
antisymmetric rotation tensor $M_W^{\scriptscriptstyle-}$ are (km
s$^{-1}$ kpc$^{-1}$)
$$ \displaylines{\hfill
M_W=\pmatrix
 { 1.3_{(0.4)}& 3.6_{(0.3)}\cr
  -1.6_{(0.3)}& 0          \cr},\hfill\llap(16)\cr\hfill
~~M_W^{\scriptscriptstyle+}=
 \pmatrix
  {1.3_{(0.4)}&1.0_{(0.2)}\cr
   1.0_{(0.2)}&0\cr},\hfill\llap(17)\cr\hfill
M_W^{\scriptscriptstyle-}=
 \pmatrix
  {0& -2.6_{(0.2)}\cr
   -2.6_{(0.2)}&0\cr}.\hfill\llap(18)
}
$$
 The deformation tensor  $M_W^{\scriptscriptstyle+}$ in
the principal axes is (km s$^{-1}$ kpc$^{-1}$):
$$
\displaylines{\hfill
 M_W^{\scriptscriptstyle+}=
 \pmatrix
 { \lambda_1&0\cr
     0&\lambda_2\cr}=\hfill\cr\hfill
=
 \pmatrix
 { 1.8&0\cr
     0&-0.5\cr},\hfill
     }
$$
and the angle between the positive direction of the $0y$ axis and
the first principal axis of this ellipse is $29\pm6^\circ$.

{
\begin{table}[p]                                                
\caption[]{\small\baselineskip=1.0ex\protect
 Kinematic parameters of the Ogorodnikov–Milne model found
by simultaneously solving the system of equations (6) and (7)
 }
\begin{center}
\begin{tabular}{|r|r|r|c|}\hline
            Parameter   &  Without correction  & With correction    \\\hline

       $X_\odot$ & $  7.99\pm0.10$ & $  7.99\pm0.10$ \\
       $Y_\odot$ & $ 16.40\pm0.10$ & $ 16.40\pm0.10$ \\
       $Z_\odot$ & $  6.72\pm0.09$ & $  6.72\pm0.09$ \\

  $A=M_{21}^{+}$ & $ 15.82\pm0.21$ & $ 15.82\pm0.21$ \\
    $M_{32}^{-}$ & $ -1.40\pm0.18$ & $ -2.60\pm0.18$ \\
    $M_{13}^{-}$ & $ -1.99\pm0.18$ & $ -0.15\pm0.18$ \\
  $B=M_{21}^{-}$ & $-11.99\pm0.15$ & $-10.87\pm0.15$ \\

 $M_{11-22}^{+}$ & $ -7.75\pm0.39$ & $ -7.75\pm0.39$ \\
    $M_{13}^{+}$ & $ -0.47\pm0.23$ & $ -0.47\pm0.23$ \\
    $M_{23}^{+}$ & $  1.00\pm0.22$ & $  1.00\pm0.22$ \\
 $M_{33-22}^{+}$ & $ -1.26\pm0.44$ & $ -1.26\pm0.44$ \\\hline

\end{tabular}
\end{center}
 \small\baselineskip=1.0ex\protect
 Note: $X_\odot,Y_\odot,Z_\odot$ in km s$^{-1}$,
 other parameters in km s$^{-1}$ kpc$^{-1}$.
 \vskip6mm
\end{table}
}
{
\begin{table}[p]                                                
\caption[]{\small\baselineskip=1.0ex\protect
Kinematic parameters
of the Ogorodnikov–Milne model found by separately solving Eqs.
(8) and (7)
 }
\begin{center}\small
\begin{tabular}{|r|r|r|r|r|}\hline
     Parameter   &   Without correction$^a$  & With correction$^a$    &  Without
 correction$^b$  & With correction$^b$ \\\hline

       $X_\odot$ & $  7.89\pm0.12$ & $  7.89\pm0.12$& $  8.51\pm0.21$ & $  8.51\pm0.21$ \\
       $Y_\odot$ & $ 16.13\pm0.13$ & $ 16.13\pm0.13$& $ 17.40\pm0.21$ & $ 17.40\pm0.21$ \\
       $Z_\odot$ &      ---        &      ---       & $  6.73\pm0.08$ & $  6.73\pm0.08$ \\

  $A=M_{21}^{+}$ & $ 15.76\pm0.25$ & $ 15.76\pm0.25$& $ 15.99\pm0.48$ & $ 15.99\pm0.48$ \\
    $M_{32}^{-}$ &      ---        &      ---       & $ -1.47\pm0.25$ & $ -2.66\pm0.25$ \\
    $M_{13}^{-}$ &      ---        &      ---       & $ -1.46\pm0.25$ & $  0.38\pm0.25$ \\
  $B=M_{21}^{-}$ & $-12.02\pm0.17$ & $-10.91\pm0.17$&      ---        &      ---        \\

 $M_{11-22}^{+}$ & $ -8.21\pm0.46$ & $ -8.21\pm0.46$& $ -4.31\pm1.03$ & $ -4.31\pm1.03$ \\
    $M_{13}^{+}$ &      ---        &      ---       & $  0.17\pm0.34$ & $  0.17\pm0.34$ \\
    $M_{23}^{+}$ &      ---        &      ---       & $  1.11\pm0.32$ & $  1.11\pm0.32$ \\
 $M_{33-22}^{+}$ &      ---        &      ---       & $  0.29\pm0.59$ & $  0.29\pm0.59$ \\

        $M_{23}$ & $  2.24\pm0.46$ & $  3.43\pm0.46$&       ---  &   --- \\
        $M_{13}$ & $ -2.93\pm0.43$ & $ -1.09\pm0.43$&       ---  &   --- \\\hline

\end{tabular}
\end{center}
 \small\baselineskip=1.0ex\protect
 $^a$The parameters were found by solving Eq. (8).

 $^b$The parameters were found by solving Eq. (7).

 Note: $X_\odot,Y_\odot,Z_\odot$ in km s$^{-1}$,
 other parameters in km s$^{-1}$ kpc$^{-1}$.

 \vskip6mm
\end{table}
}

Equation (1) can be written as
$$ \displaylines{\hfill
 {\bf V}={\bf V}_\circ+{\rm grad}~F +
 ({\hbox {\boldmath $\omega$}}\times{\bf r}).\hfill
 }
$$
Analysis of grad~$F$ shows (Sedov 1970) that an infinitesimal
sphere of radius $r$ composed of points of the medium at time $t$,
$$
\displaylines{\hfill x^2+y^2+z^2=r^2 \hfill}
$$
transforms into a deformation ellipsoid after time $\Delta t$,
$$ \displaylines{\hfill
 {x^{*2}\over{(1+\lambda_1 \Delta t)^2}}+
 {y^{*2}\over{(1+\lambda_2 \Delta t)^2}}+
 {z^{*2}\over{(1+\lambda_3 \Delta t)^2}}=r^2, \hfill
 }
$$
which is shown in Fig. 2b for our two-dimensional case.

For case 2, we have $M_{\scriptscriptstyle33}=-1.3\pm0.4$~km
s$^{-1}$ kpc$^{-1}$ and $M_{\scriptscriptstyle22}=0$. The
deformation tensor $M_W^{\scriptscriptstyle+}$ in the principal
axes is (km s$^{-1}$ kpc$^{-1}$)
$$ \displaylines{
M_W^{\scriptscriptstyle+}=
 \pmatrix
 { 0.5&0\cr
     0&-1.8\cr},
}
$$
 the angle
between the positive direction of the $0y$ axis and the first
principal axis of this ellipse is $29\pm6^\circ$.

For case 3, where $M_{\scriptscriptstyle33}=0$ and
$M_{\scriptscriptstyle22}=0$, the deformation tensor
$M_W^{\scriptscriptstyle+}$ has two roots: $\lambda_1=1.0$~km
s$^{-1}$ kpc$^{-1}$ and $\lambda_2=-1.0$~km s$^{-1}$ kpc$^{-1}$;
therefore, the first principal axis of this ellipse is oriented at
an angle of $45^\circ$ to the $0y$ axis.

In all three cases, the divergence
$0.5(M_{\scriptscriptstyle22}+M_{\scriptscriptstyle33})$ is
insignificant. It is $+0.6\pm0.4$, and 0 km s$^{-1}$ kpc$^{-1}$
for the first, second, and third cases, respectively.

As a result, we can conclude that the kinematics in the $yz$ plane
can be described as a rotation around the Galactic $x$ axis with
an angular velocity
$\Omega_W=M_{\scriptscriptstyle32}^{\scriptscriptstyle-}-\lambda_1$.
This angular velocity is $-4.4\pm0.5$, $-3.1\pm0.5$, and
$-3.6\pm0.3$~km s$^{-1}$ kpc$^{-1}$ for the first, second, and
third cases, respectively.

\begin{figure}[p]{
\begin{center}
 \includegraphics[width=80mm]{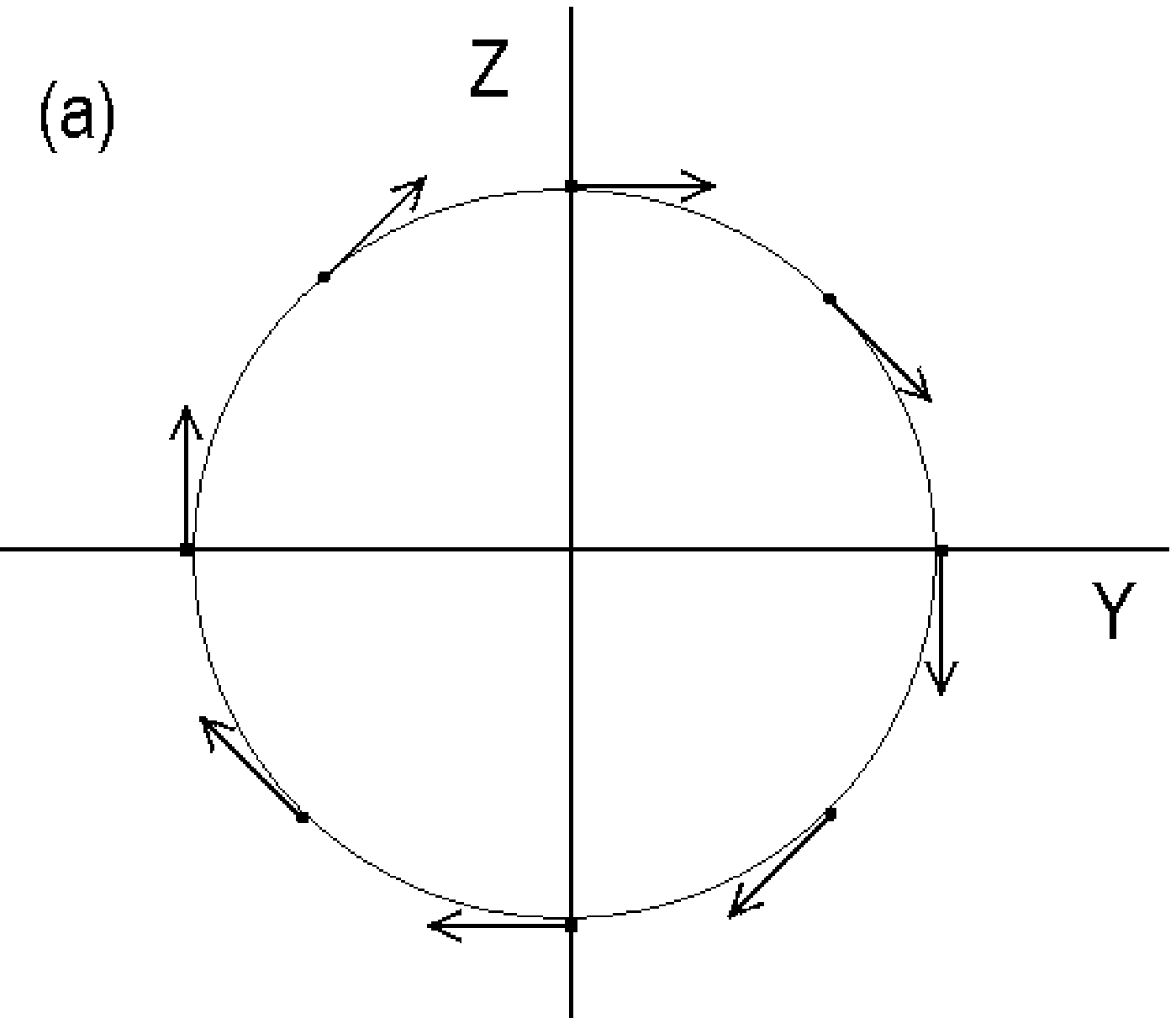}

 \includegraphics[width=80mm]{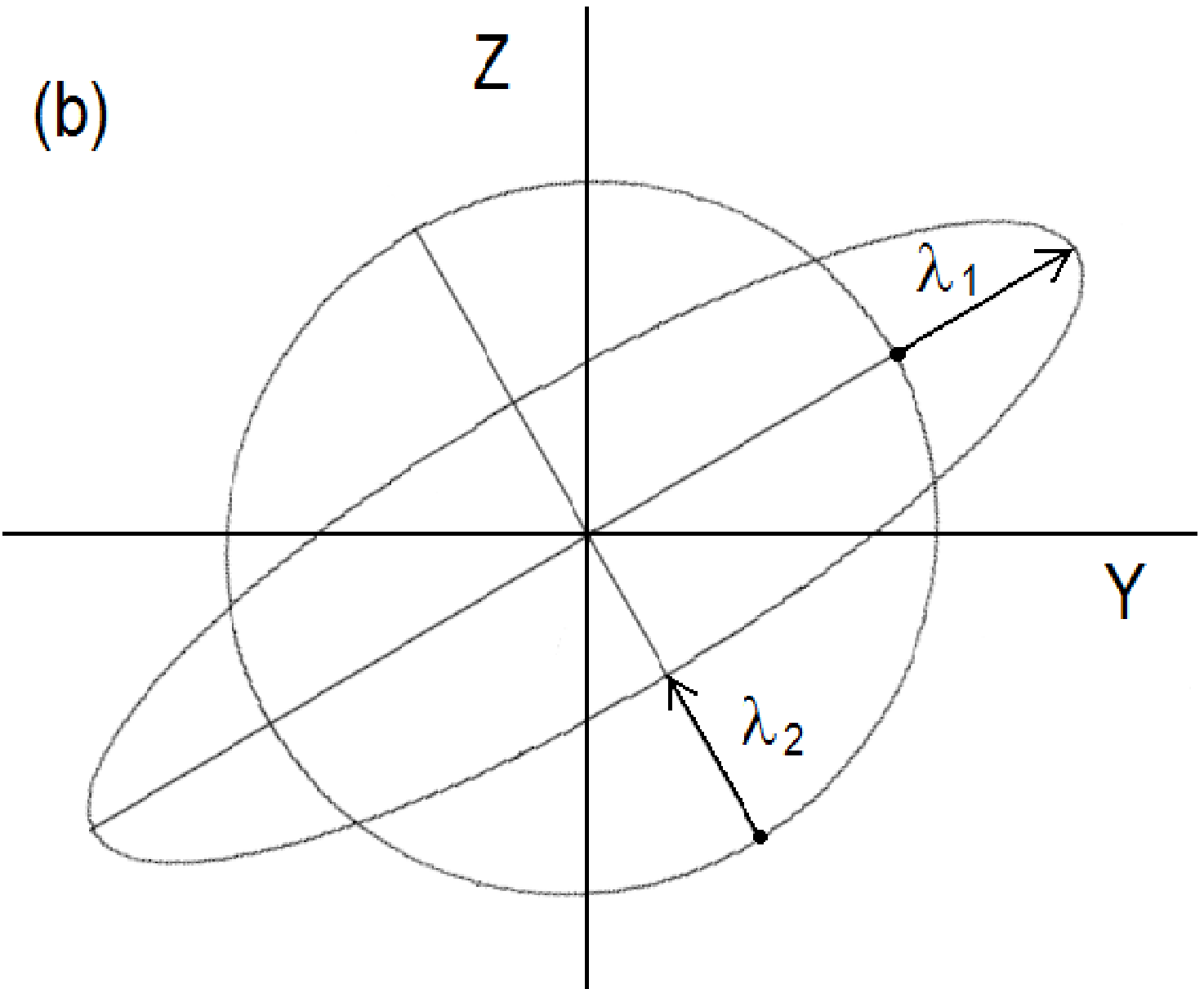}
\vskip5mm
\end{center}}
Fig. 2. Distribution of rotation velocity vectors (a) and
deformation velocity vectors (b) in the $yz$ plane.
\end{figure}

Let us estimate the linear velocities in the $yz$ plane. The mean
photometric distance for our sample of RGC stars is ${\overline
r}=0.57\pm0.17$~kpc. The linear solidbody rotation velocity is
then $M_{\scriptscriptstyle32}^{\scriptscriptstyle-} \cdot
{\overline r}=-1.5\pm0.1$~km s$^{-1}$. The maximum deformation
velocity is $\lambda_1 \cdot {\overline r}= 1.0\pm0.3$~km s$^{-1}$
and the highest (among the cases considered) linear velocity is
$\Omega_W \cdot {\overline r}=-2.5\pm0.3$~km s$^{-1}$.

Figure 2 gives a qualitative picture that reflects the pattern of
our results. The distribution of rotation velocity vectors in the
$yz$ plane for the case of rotation we found is shown in Fig. 2a.
Figure 2b shows how a circumference of unit radius turns into a
deformation ellipse after a unit time interval.

According to the results of our separate solution of Eqs. (8) and
(7) (the third and fifth columns of Table 3), for example, for
case 1, we have
$$ \displaylines{\hfill
M_W=\pmatrix
 { -0.3_{(0.6)}& 3.8_{(0.5)}\cr
  -1.6_{(0.4)}& 0          \cr},\hfill\llap(19)\cr\hfill
~~M_W^{\scriptscriptstyle+}=
 \pmatrix
  {-0.3_{(0.6)}&1.1_{(0.3)}\cr
   1.1_{(0.3)}&0\cr},\hfill\llap(20)\cr\hfill
M_W^{\scriptscriptstyle-}=
 \pmatrix
  {0& -2.7_{(0.3)}\cr
   -2.7_{(0.3)}&0\cr}.\hfill\llap(21)
}
$$
We can see from our comparison of matrixes (19) and (16), (20) and
(17), (21) and (18) that there are differences only in the
diagonal element. This leads us to conclude that both approaches
used (the simultaneous and separate solution of the equations) are
equivalent, because they yield coincident results.

When writing Eq. (14), we set the components equal to zero,
$\omega_x=0$ and $\omega_y=0$. As we see from the Mean 2 solution
in Table 1, the value of the component $\omega_y$ was found to be
nonzero outside the $2\sigma$ range. However, the kinematic
parameters calculated by taking this into account, i.e.,
$\omega_y\neq0$ and $\omega_z\neq0$, have no significant
differences between the solutions reflected in Tables 2 and 3 and,
hence, we do not give them.

\subsection*{DISCUSSION}

Drimmel et al. (2000) provide arguments for the model of
precession of the warped disk in the $zx$ plane (rotation around
the y axis) with an angular velocity of $-25$~km s$^{-1}$
kpc$^{-1}$. However, as can be seen from Table 2, as a result of
applying the correction $\omega_z=-0.52$ mas yr$^{-1}$, the
component of the rotation tensor around the $y$ axis
($M_{\scriptscriptstyle13}^{\scriptscriptstyle-}$) has an almost
zero value and the component of the deformation tensor in this
plane ($M_{\scriptscriptstyle13}^{\scriptscriptstyle+}$) does not
differ significantly from zero.

The angular velocity $\Omega_W$ estimated here is inconsistent
with the analysis of the motion of giants of various spectral
types (O--M) performed by Miyamoto et al. (1993), who found a
positive direction of rotation around the $x$ axis (directed
toward the Galactic center). Note that the analyzed stellar proper
motions were determined in the FK5 (Fundamental Catalog 5) system,
which is noticeably distorted by the uncertainty in the precession
constant. On the other hand, having analyzed the proper motions of
O--B5 stars in the HIPPARCOS system,Miyamoto and Zhu (1998) also
reached the conclusion about a positive rotation around the $x$
axis.

At present, it is impossible to choose between the three forms of
the deformation tensor considered in the previous section due to
the absence of data on stellar radial velocities and distances.
Note that based on a sample of 3632 relatively close RGC stars
with known HIPPARCOS trigonometric parallaxes and radial
velocities, we estimated the parameters
$M_{\scriptscriptstyle22}=-1.3\pm3.2$~km s$^{-1}$ kpc$^{-1}$ and
 $M_{\scriptscriptstyle33}=-1.1\pm2.6$~km s$^{-1}$
kpc$^{-1}$ (Bobylev et al. 2009). Consequently, the difference
$(M_{\scriptscriptstyle33}^{\scriptscriptstyle+}-
 M_{\scriptscriptstyle22}^{\scriptscriptstyle+})$ is very close to zero and, hence, the orientation of the
deformation ellipse is close to $45^\circ$, as in case 3
considered.

In the Introduction, we listed various hypotheses about the nature
of the Galactic disk warp. In our opinion, the hypothesis about an
outflow of gas from the Magellanic Clouds (Olano 2004) has been
worked out in greatest detail. Assuming the distance to the Large
Magellanic Cloud to be $r_{LMC}=50$~kpc, we find that the linear
velocity of the rotation we revealed at this distance,
$(-3.1\pm0.5)\leq\Omega_W\leq(-4.4\pm0.5)$ km s$^{-1}$ kpc$^{-1}$,
is $|\Omega_W| \cdot r_{LMC}=(155\div220)\pm25$~km s$^{-1}$. Such
velocities are in agreement with the estimate of the motion of
high-velocity hydrogen clouds relative to the local standard of
rest, $\approx200$~km s$^{-1}$ (Olano 2004), and the rotation
direction we found is in good agreement with the direction of
motion of the Magellanic Clouds around the Galaxy during the past
570 Myr (Fig. 6 from Olano 2004).

\subsection*{CONCLUSIONS}

We analyzed the kinematics of about 82000 RGC stars from the
Tycho-2 catalog. These stars are a variety of ``standard
candles'', because they occupy a compact region on the
Hertzsprung-Russell diagram. Therefore, photometric distance
estimates with an accuracy of at least 30\% are available for
them. The RGC stars considered lie in the range of heliocentric
distances 0.3 kpc $<d<1$ kpc.


Since these stars are distributed fairly uniformly in space, they
are of great interest not only in studying the Galaxy’s general
rotation described by the rotation parameters in the Galactic $xy$
plane but also in the other two planes, namely $yz$ and $zx$.

For a reliable description of their kinematic peculiarities, we
should be confident that the observational data (Tycho-2 proper
motions) are free from the systematic errors related to the
referencing of the optical realization of the ICRS/HIPPARCOS
system to the inertial frame of reference specified by
extragalactic sources.

Here, we gave much attention to this problem. Based on all of the
currently available data, we determined new, most probable
components of the residual rotation vector of the optical
realization of the ICRS/HIPPARCOS system relative  to the inertial
frame of reference,
$(\omega_x,\omega_y,\omega_z)=(-0.11,0.24,-0.52)\pm(0.14,0.10,0.16)$
mas yr$^{-1}$. This led us to conclude that a small correction,
$\omega_z=-0.52$ mas yr$^{-1}$, should be applied to the Tycho-2
stellar proper motions of the form $\mu_\alpha\cos\delta$.

By applying this correction, we showed that, apart from their
involvement in the general Galactic rotation described by the Oort
constants $A=15.8\pm0.2$~km s$^{-1}$ kpc$^{-1}$ and
$B=-10.9\pm0.2$~km s$^{-1}$ kpc$^{-1}$, the RGC stars considered
have peculiarities in the $yz$ plane related, in our opinion, to
the kinematics of the warped Galactic stellar-gaseous disk.

The component of the solid-body rotation of the local solar
neighborhood around the Galactic x axis is
$M_{\scriptscriptstyle32}^{\scriptscriptstyle-}=-2.6\pm0.2$ km
s$^{-1}$ kpc$^{-1}$. The parameter of the deformation tensor in
this plane
$M_{\scriptscriptstyle23}^{\scriptscriptstyle+}=1.0\pm0.2$ km
s$^{-1}$ kpc$^{-1}$ and the difference
$M_{\scriptscriptstyle33}-M_{\scriptscriptstyle22}=-1.3\pm0.4$ km
s$^{-1}$ kpc$^{-1}$ differ significantly from zero.

On the whole, the kinematics of the warped Galactic
stellar–gaseous disk in the local solar neighborhood can be
described as a rotation around the Galactic x axis directed from
the Sun toward the Galactic center with an angular velocity
$(-3.1\pm0.5)\leq\Omega_W\leq(-4.4\pm0.5)$~km s$^{-1}$ kpc$^{-1}$.
Thus, the rotation is around an axis close to the line of nodes of
this structure. In the case where the difference
$M_{\scriptscriptstyle33}-M_{\scriptscriptstyle22}$ may be set
equal to zero, the angular velocity $\Omega_W$ is $-3.6\pm0.3$~km
s$^{-1}$ kpc$^{-1}$. If $M_{\scriptscriptstyle22}=0$ and
$M_{\scriptscriptstyle33}\neq0$, then $\Omega_W=-3.1\pm0.5$~km
s$^{-1}$ kpc$^{-1}$. If $M_{\scriptscriptstyle22}\neq0$ and
$M_{\scriptscriptstyle33}=0$, then $\Omega_W$ reaches
$-4.4\pm0.5$~km s$^{-1}$ kpc$^{-1}$.

 \bigskip
{\bf ACKNOWLEDGMENTS}
 \bigskip

I am grateful to the referees for helpful remarks that contributed
to an improvement of the paper and to A.T. Bajkova for her help in
the work. This study was supported by the Russian Foundation for
Basic Research (project nos. 08--02--00400 and
09--02--90443--Ukr\_f) and, in part, by the ``Origin and Evolution
of Stars and Galaxies'' Program of the Presidium of the Russian
Academy of Sciences.

 \bigskip
{\bf REFERENCES}
 \bigskip

1. J. Bailin, Astrophys. J. 583, L79 (2003).

2. E. Battaner, E. Florido, and M.L. Sanchez-Saavedra, Astron.
Astrophys. 236, 1 (1990).

3. D.A. Boboltz, A.L. Fey, W.K. Puatua, et al., Astron. J. 133,
906 (2007).

4. V.V. Bobylev, Pis’ma Astron. Zh. 30, 289 (2004a) [Astron. Lett.
30, 251 (2004)].

5. V.V. Bobylev, Pis’ma Astron. Zh. 30, 930 (2004b) [Astron. Lett.
30, 848 (2004)].

6. V.V. Bobylev, N.M. Bronnikova, and N.A. Shakht, Pis’ma Astron.
Zh. 30, 519 (2004) [Astron. Lett. 30, 469 (2004)].

7. V.V. Bobylev, A.S. Stepanishchev, A.T. Bajkova, and G.A.
Gontcharov, Pis’ma Astron. Zh. 35, 920 (2009) [Astron. Lett. 35,
836 (2009)].

8. V.V. Bobylev, P.N. Fedorov, A.T. Bajkova, and V.S. Akhmetov,
Pis’ma Astron. Zh. 36, 535 (2010) [Astron. Lett. 36, 506 (2010)].

9. W.B. Burton, Galactic and Extragalactic Radio Astronomy, Ed. by
G. Verschuur and K. Kellerman (Springer, New York, 1988), p. 295.

10. J.C. Cersosimo, S. Mader, N. Santiago Figueroa, et al.,
Astrophys. J. 699, 469 (2009).

11. Yu.A. Chernetenko, Pis’ma Astron. Zh. 34, 296 (2008) [Astron.
Lett. 34, 266 (2008)].

12. S.V.M. Clube, Mon. Not. R. Astron. Soc. 159, 289 (1972).

13. S.V.M. Clube, Mon. Not. R. Astron. Soc. 161, 445 (1973).

14. W. Dehnen, Astron. J. 115, 2384 (1998).

15. R. Drimmel and D.N. Spergel, Astrophys. J. 556, 181 (2001).

16. R. Drimmel, R.L. Smart, and M.G. Lattanzi, Astron. Astrophys.
354, 67 (2000).

17. P.N. Fedorov, A.A. Myznikov, and V.S. Akhmetov, Mon. Not. R.
Astron. Soc. 393, 133 (2009).

18. M. Geffert, A.R. Klemola, M. Hiesgen, et al., Astron.
Astrophys. 124, 157 (1997).

19. G.A. Gontcharov, Pis’ma Astron. Zh. 34, 868 (2008) [Astron.
Lett. 34, 785 (2008)].

20. P.D. Hemenway, R.L. Duncombe, E.P. Bozyan, et al., Astron. J.
114, 2796 (1997).

21. The HIPPARCOS and Tycho Catalogues, ESA SP-1200 (1997).

22. S. Hirte, E. Schilbach, and R.-D. Scholz, Astron. Astrophys.
Suppl. Ser. 126, 31 (1996).

23. E. Hog, C. Fabricius, V.V. Makarov, et al., Astron. Astrophys.
355, L27 (2000).

24. P.M.W. Kalberla and L. Dedes, Astron. Astrophys. 487, 951
(2008).

25. V.S. Kislyuk, SP. Rybka, A.I. Yatsenko, et al., Astron.
Astrophys. 321, 660 (1997).

26. A.R. Klemola, R.B. Hanson, and B.F. Jones, in Galactic and
Solar System Optical Astrometry, Ed. by L.V. Morrison and G.F.
Gilmore (Cambridge Univ., Cambridge, 1994), p. 20.

27. J. Kovalevsky, L. Lindegren, M.A.C. Perryman, et al., Astron.
Astrophys. 323, 620 (1997).

28. L. Lindegren and J. Kovalevsky, Astron. Astrophys. 304, 189
(1995).

29. M. L\'opez-Corredoira, J. Betancort-Rijo, and J. Beckman,
Astron. Astrophys. 386, 169 (2002).

30. C. Ma, E.F. Arias, T.M. Eubanks, et al., Astron. J. 116, 516
(1998).

31. M. Miyamoto and Z. Zhu, Astron. J. 115, 1483 (1998).

32. M. Miyamoto, M. S\^oma, and M. Yoshizawa, Astron. J. 105, 2138
(1993).

33. Y. Momany, S. Zaggia, G. Gilmore, et al., Astron. Astrophys.
451, 515 (2006).

34. D.G.Monet, Bull. Am. Astron. Soc. 30, 1427 (1998).

35. K.F. Ogorodnikov, Dynamics of Stellar Systems (Fizmatgiz,
Moscow, 1965) [in Russian].

36. C.A. Olano, Astron. Astrophys. 423, 895 (2004).

37. I. Platais, V. Kozhurina-Platais, T.M. Girard, et al., Astron.
Astrophys. 331, 1119 (1998a).

38. I. Platais, T.M. Girard, V. Kozhurina-Platais, et al., Astron.
J. 116, 2556 (1998b).

39. S.P. Rybka, Kinem Fiz. Neb. Tel 22, 225 (2006).

40. S.P. Rybka and A.I. Yatsenko, Astron. Astrophys. Suppl. Ser.
121, 243 (1997).

41. L.I. Sedov, Mechanics of Continuous Media (Nauka, Moscow,
1970), Vol. 1 [in Russian].

42. M.F. Skrutskie, R.M. Cutri, R. Stiening, et al., Astron. J.
131, 1163 (2006).

43. L. Sparke and S. Casertano, Mon. Not. R. Astron. Soc. 234, 873
(1988).

44. H.-J. Tucholke, P. Brosche, and M. Odenkirchen, Astron.
Astrophys. Suppl. Ser. 124, 157 (1997).

45. V.V. Vityazev and A.S. Tsvetkov, Pis’ma Astron. Zh. 35, 114
(2009) [Astron. Lett. 35, 100 (2009)].

46. J. Vondr\'ak, C.~Ron, I.~Pe\v sek, Astron. Astrophys. 319,
1020 (1997).

47. G. Westerhout, Bull. Astron. Inst. Netherlands 13, 201 (1957).

48. S.K. Yi, Y.-C. Kim, and P. Demarque, Astrophys. J. Suppl. Ser.
144, 259 (2003).

49. I. Yusifov, astro-ph/0405517 (2004).

50. N. Zacharias, S.E. Urban, M.I. Zacharias, et al., Astron. J.
127, 3043 (2004).

51. Zi Zhu, Publ. Astron. Soc. Jpn. 53, L33 (2001).

52. Zi Zhu, in JOURNE\'ES-2003, Astrometry, Geodynamics and Solar
System Dynamics: From Milliarcseconds to Microarcseconds, Ed. by
N. Capitaine (Observ. de Paris, Paris, 2003), p. 95.

\bigskip
Translated by
 N. Samus’

\end{document}